\documentclass[showpacs,preprintnumbers]{revtex4}
\usepackage{amsmath}
\usepackage{graphicx}

\begin{document}

\textbf{Comment on \textquotedblleft Trouble with the Lorentz Law of Force:
Incompatibility with Special}

\textbf{Relativity and Momentum Conservation\textquotedblright \bigskip }

Tomislav Ivezi\'{c}

\textit{Ru%
\mbox
{\it{d}\hspace{-.15em}\rule[1.25ex]{.2em}{.04ex}\hspace{-.05em}}er Bo\v{s}%
kovi\'{c} Institute, P.O.B. 180, 10002 Zagreb, Croatia}

\textit{ivezic@irb.hr\bigskip }

In [1] it is argued that in the presence of the magnetization $\mathbf{M}$
and the electric polarization $\mathbf{P}$ the usual expression for the
Lorentz force with three-dimensional (3D) vectors leads to an apparent
paradox; in a static electric field $\mathbf{E}$ a magnetic dipole moment
(MDM) $\mathbf{m}$ is subject to a torque $\mathbf{T}$ in some frames and
not in others. In [1, 2] it is concluded that the Lorentz force should be
replaced by the Einstein-Laub law, which predicts no torque $\mathbf{T}$ in
all frames. Note that in [1, 2] all quantities $\mathbf{E}$, $\mathbf{B}$, $%
\mathbf{P}$, $\mathbf{T}$, etc. are the 3D vectors and it is considered that
their transformations (the usual transformations (UT)) are the
relativistically correct Lorentz transformations (LT)\ (boosts). In [3] the
\textquotedblleft resolutions\textquotedblright\ of the paradox from [1] are
presented taking into account some \textquotedblleft
hidden\textquotedblright\ 3D quantities. Here, we show that the principle of
relativity is naturally satisfied and there is no paradox if an independent
physical reality is attributed to the 4D geometric quantities (GQs) and not,
as usual, to the 3D quantities. Hence, there is no need either for the
change of the expression for the Lorentz force, but as a 4D GQ, or for the
introduction of some \textquotedblleft hidden\textquotedblright\ 3D
quantities. All this is already presented in detail in [4, 5] and here I
summarize some results from [4, 5].

For the UT of the 3D $\mathbf{E}$, $\mathbf{B}$ see Sec. 3.1 and for $%
\mathbf{P}$, $\mathbf{M}$ Sec. 3.2 in [4], see also Eqs. (11-12b) and Eqs.
(9-10b) in [1].\ It is explained in Secs. 5 and 6 in [4] or Sec. 3 in [5]
that, in the 4D spacetime, the UT of the 3D fields \emph{are not} the LT. In
contrast to the UT, \emph{the LT always transform the 4D algebraic object
representing, e.g., the electric field only to the electric field; there is
no mixing with the magnetic field. }For example, the LT of the components $%
E^{\mu }$ (in the standard, $\{\gamma _{\mu }\}$ basis) of the electric
field vector $E=E^{\mu }\gamma _{\mu }$ are given as $E^{0}=\gamma
(E^{\prime 0}+\beta E^{\prime 1})$, $E^{1}=\gamma (E^{\prime 1}+\beta
E^{\prime 0})$, $E^{2,3}=E^{\prime 2,3}$, for a boost along the $x^{1}$
axis. For a short derivation of these LT see, e.g., [6]. In the 4D spacetime
the vector $E$ \emph{is the same 4D quantity} for all inertial observers,
i.e., it holds $E=E^{\nu }\gamma _{\nu }=E^{\prime \nu }\gamma _{\nu
}^{\prime }=...$ . The same LT hold for any other vector, e.g., $x$, $B$, $P$%
, $M$, EDM $p$ and MDM $m$, etc. In [4, 5] it is also shown that neither the
\textquotedblleft resolution\textquotedblright\ of the mentioned paradox by
means of the Einstein-Laub law [1, 2] and also all other \textquotedblleft
resolutions\textquotedblright\ from [3] are relativistically correct since
they all deal with the 3D vectors and their UT. Hence, they are
synchronization dependent, e.g., they are meaningless for a nonstandard
\textquotedblleft radio\textquotedblright\ synchronization, see Sec. 3.1 in
[4]. But, \emph{every synchronization is only a convention and physics must
not depend on conventions.}

Using the 4D GQs it is shown, e.g., in Sec. 8 in [4] that a \emph{stationary}
permanent magnet possesses not only an intrinsic magnetization $M$ but also
an intrinsic electric polarization $P$. That result was derived using the
generalized Uhlenbeck-Goudsmit hypothesis [7] according to which the
connection between the dipole moment bivector $D$ and the spin bivector $%
\mathcal{S}$ is given as $D=g_{S}S$, Eq. (9) in [7]. Hence, \emph{in a
static electric field, both, a current-loop and a permanent magnet
experience the Lorentz force }(vector) $K_{L}$ \emph{and the torque}
(bivector) $N$ \emph{in all relatively moving inertial frames and there is
no paradox}. Thus, we consider that the \emph{whole} bivector $N$ is
correctly defined quantity in the 4D spacetime and not the usual 3D torque $%
\mathbf{T}=\mathbf{r}\times \mathbf{F}$. \emph{In the 4D spacetime there is
no room for the 3-vectors; they cannot correctly transform under the LT.} In
the magnet's rest frame $S^{\prime }$ the 4D torque $N$ is given by Eqs.
(71-73) in [4]; $N=-(1/c)E^{\prime 1}m^{\prime 2}(\gamma _{0}^{\prime
}\wedge \gamma _{3}^{\prime })-E^{\prime 1}p^{\prime 3}(\gamma _{1}^{\prime
}\wedge \gamma _{3}^{\prime })$, where, in the considered case, $E=E^{\prime
1}\gamma _{1}^{\prime }$, $m=m^{\prime 2}\gamma _{2}^{\prime }$ and $%
p=p^{\prime 3}\gamma _{3}^{\prime }$. In $S$, the lab frame, the torque $N$
can be obtained by the correct LT of the 4D GQs from $S^{\prime }$ and it is
$N=(-E^{1}m^{2}/c+\beta E^{1}p^{3})(\gamma _{0}\wedge \gamma _{3})+(\beta
E^{1}m^{2}/c-E^{1}p^{3})(\gamma _{1}\wedge \gamma _{3})$. \emph{The 4D torque%
} $N$ \emph{is the same 4D GQ for all relatively moving inertial observers},
$N=(1/2)N^{^{\prime }\mu \nu }\gamma _{\mu }^{\prime }\wedge \gamma _{\nu
}^{\prime }=(1/2)N^{\mu \nu }\gamma _{\mu }\wedge \gamma _{\nu }$, \emph{and
there is no paradox.} It can be seen from these equations for $N$ that it
will be the same quantity in $S^{\prime }$ and $S$\ even in the case that $%
p=0$ and again there is no paradox, see the end of Sec. 4 in [5]. Note that
the same objections as for [1-3] hold in the same measure for all other
\textquotedblleft resolutions\textquotedblright\ from Ref. [4] in
[5].\bigskip

\noindent \textbf{References\bigskip }

\noindent \lbrack 1] M. Mansuripur, Phys. Rev. Lett. \textbf{108}, 193901
(2012).

\noindent \lbrack 2] M. Mansuripur, Phys. Rev. Lett. \textbf{110}, 089405
(2013); Proc. SPIE \textbf{8455}, 845512 (2012).

\noindent \lbrack 3] D. A. T. Vanzella, Phys. Rev. Lett. \textbf{110},
089401 (2013); S. M. Barnett, Phys. Rev. Lett. \textbf{110},

089402 (2013); P. L. Saldanha, Phys. Rev. Lett. \textbf{110}, 089403 (2013);
M. Khorrami,

Phys. Rev. Lett. \textbf{110}, 089404 (2013).

\noindent \lbrack 4] T. Ivezi\'{c}, J. Phys.: Conf. Ser. \textbf{437},
012014 (2013); arXiv: 1204.5137.

\noindent \lbrack 5] T. Ivezi\'{c}, arXiv: 1212.4684.

\noindent \lbrack 6] T. Ivezi\'{c}, Phys. Rev. Lett. \textbf{98}, 108901
(2007).

\noindent \lbrack 7] T. Ivezi\'{c}, Phys. Scr.\textit{\ }\textbf{81}, 025001
(2010).

\end{document}